\documentclass[%
 reprint,
superscriptaddress,
 amsmath,amssymb,
 aps,
]{revtex4-2}

\usepackage{graphicx}
\usepackage{dcolumn}
\usepackage{bm}
\usepackage{color}
\usepackage{hyperref}

\begin{document}

\title{Quantum phase transition in two-dimensional NbN\\
 superconducting thin films}

\author{Tian-Yu Jing}
\affiliation{Tianjin Key Laboratory of Low Dimensional Materials Physics and
Preparing Technology, Department of Physics, Tianjin University, Tianjin 300072,
China}
\author{Zi-Yan Han}
\affiliation{Tianjin Key Laboratory of Low Dimensional Materials Physics and
Preparing Technology, Department of Physics, Tianjin University, Tianjin 300072,
China}
\author{Zhi-Hao He}
\affiliation{Tianjin Key Laboratory of Low Dimensional Materials Physics and
Preparing Technology, Department of Physics, Tianjin University, Tianjin 300072,
China}
\author{Ming-Xin Shao}
\affiliation{State Key Laboratory of Electronic Thin Films and Integrated Devices, University of Electronic Science and Technology of China, Chengdu 611130, China}

\author{Peng Li}
\affiliation{State Key Laboratory of Electronic Thin Films and Integrated Devices, University of Electronic Science and Technology of China, Chengdu 611130, China}
\author{Zhi-Qing Li}
\email[Corresponding author, e-mail: ]{zhiqingli@tju.edu.cn}
\affiliation{Tianjin Key Laboratory of Low Dimensional Materials Physics and
Preparing Technology, Department of Physics, Tianjin University, Tianjin 300072,
China}
\date{\today}

\begin{abstract}
We systematically investigated the low-temperature transport properties of a series of NbN epitaxial films with thickness $t$ ranging from $\sim$2.0 to $\sim$4.0\,nm. The films undergo a superconductor-insulator transition (SIT) with decreasing film thickness, and the critical sheet resistance for the SIT is close to the quantum resistance of Cooper pairs $h/4e^2$ (6.45\,k$\Omega$). Besides the Berezinski-Koterlitz-Thouless transition, a magnetic-field-driven SIT is observed in those two-dimensional (2D) superconducting films ($2.6\,{\rm nm}\lesssim t \lesssim 4.0\,{\rm nm}$). Interestingly, it is found that the low-temperature magnetoresistance isotherms do not cross at a single fixed point but at a well-distinguished region for these superconducting films. The dynamical critical exponent obtained by analyzing these magnetoresistance isotherms is divergent as the quantum critical point is being approached. The behavior of the dynamical critical exponent, originating from quenched disorder at ultralow temperatures, provides direct evidence for the occurrence of quantum Griffiths singularity in the quantum phase transition process of the films. The field-driven anomalous metal (quantum metal) state does not appear in these films. Our results suggest that the quantum Griffiths singularity not only occurs in the highly crystalline 2D superconductors with superconductor-metal transition but also in those with SIT.
\end{abstract}

\maketitle


\section{Introduction}\label{secI}
Two-dimensional (2D) superconductors are good systems for investigating quantum phase transition and have attracted great attention over the past decades. The superconductor-insulator/metal transition (SIT/SMT) in 2D superconductors is a typical quantum phase transition and can be tuned by magnetic field~\cite{PrlHebard, PrlYazdani}, carrier concentration~\cite{SciSaito}, or degree of disorder~\cite{PrlBaturina, PrbBreznay}. Generally, the SIT in 2D homogeneous superconductors is explained in the framework of ``dirty-boson model", which predicts that the ground state of the system directly changes from superconducting to insulating states without passing through an intermediate metal regime. In addition, this model predicts that a magnetic field would induce a SIT, in which the sheet resistance of the superconducting film  satisfies a power-law scaling form~\cite{PrlFisher1, PrbFisher}
\begin{equation}\label{Eq-scaling}
 R_\square(B, T)=R_\square^{\rm c} f(\delta T^{-1/\nu z}),
 \end{equation}
where $\delta=|B-B_{\rm c}|$ is the distance from the critical field $B_{\rm c}$, $R_\square^{\rm c}$ is the critical sheet resistance, $\nu$ is the correlation length exponent, $z$ is the dynamical critical exponent, and $f(x)$ is the scaling function with $f(0)=1$. The scaling form in Eq.~(\ref{Eq-scaling}) implies that the magnetic field dependence of resistance curves at different temperatures all cross at the critical field $B_{\rm c}$. Recent experimental results have indicated that an intermediate anomalous metallic state appears in some 2D superconductors, including amorphous Mo$_{43}$Ge$_{57}$~\cite{PrlEphron} and Ta thin films~\cite{PrbLi}, ZrNCl-based electric-double layer transistors~\cite{SciSaito}, mechanical exfoliated crystalline NbSe$_2$~\cite{NpTsen} and WTe$_2$ films~\cite{SciSajadi}, and Josephson junction arrays~\cite{PrbVan der Zant, PrlVan der Zant}.  More recently, it has been found that in some 2D metal superconductors (i.e., the normal states of these superconductors reveal metallic characteristics) the magnetoresistance isotherms do not cross at a fixed point (the critical field) but at multiple points, and the effective exponent $z\nu$ determined at each crossing point diverges as the quantum phase transition is approached~\cite{SciXing, PrbShen, NlXing, NcSaito, PrbLewellyn, CpZhang, SbHuang}. This phenomenon is the so-called quantum Griffiths singularity (QGS) of SMT. The appearance of QGS in 2D superconductors does not conform to the ``dirty-boson model" either.

The QGS of SMT was initially observed in 3-monolayer Ga film~\cite{SciXing}, and subsequently observed in some highly crystalline 2D superconductors with metallic normal states, such as LaAlO$_3$/SrTiO$_3$ interface~\cite{PrbShen}, macro-size monolayer NbSe$_2$~\cite{NlXing}, and ion-gated ZrNCl and MoS$_2$~\cite{NcSaito}.  Since 2019, the QGS has also been observed in amorphous InO$_x$~\cite{PrbLewellyn}, WSi~\cite{CpZhang}, and $\beta$-W~\cite{SbHuang} films. It is believed that quenched disorder plays an important role in the occurrence of QGS in highly crystalline 2D superconductors~\cite{SciXing, NlXing, NcSaito}. According to previous reports~\cite{SciXing, NlXing, NcSaito}, those highly crystalline 2D superconductors with QGS generally reveal metallic characteristics above the superconducting transition temperature $T_{\rm c}$. Therefore, it is necessary to explore whether there is QGS in highly crystalline 2D superconducting film with insulating normal state. In addition, it is interesting to check whether the intermediate anomalous metal state presents in these superconducting films with QGS during the quantum phase transition.

The 2D NbN superconductor films could be a suitable system to address the above issues. There are several advantages for using NbN films. (1) The superconducting transition temperature of NbN film can still be as high as 9\,K even if the film thickness is only 3 nm~\cite{EplKundu}. (2) A NbN film possesses high resistance to oxidation at room temperature. (3) It has been demonstrated that supeconducting puddles (or islands) would appear in the NbN film below $T_{\rm c}$~\cite{PrbNoat, PrbCarbillet2013, PrbCarbillet2016}, which could be the necessary condition for the occurrence of QGS in 2D superconductors. In the present paper, we systematically investigate the electrical transport properties of a series of NbN films with different thickness. For the 2D superconducting films, the magnetic-field-induced SIT as well as the QGS is observed. However, the anomalous metallic state does not appear in these films. We will report these interesting phenomena in the following sections.

\section{Experimental method}
Our NbN films with thickness ranging from $\sim$2.0 to $\sim$100 nm  were grown on (100) MgO single crystal substrates by the standard reactive magnetron sputtering method.  A commercial niobium target with purity of 99.99\% was used as the sputtering source. The base pressure of the chamber was less than $1 \times 10^{-5}$\,Pa. The deposition was carried out in a mixture of argon (99.999\%) and nitrogen (99.999\%) atmosphere and the pressure of the chamber (working pressure) was kept at 0.2\,Pa. During the deposition, the sputtering power was set as 300\,W. It was found that the optimal substrate temperature and volume ratio of nitrogen to argon were 763\,K and $1:7$, respectively.

The thicknesses of the films $d$ were evaluated through growth rate and deposition time: fixing the deposition conditions, including sputtering power, substrate temperature, ratio of nitrogen to argon, and working pressure, we first fabricated some thick films ($t > 200$\, nm);  the thicknesses of these thick films were measured using a surface profiler (Dektak, 6M), and then the growth rate (12\,nm/min) was obtained. The thicknesses of the $t\lesssim 20$\,nm films were further determined by the high resolution transmission electron microscopy (HRTEM) of the cross section of the films. Crystal structure and phase characterization were determined by x-ray diffraction (XRD) using Cu $K_\alpha$ radiation at room temperature. The microstructure of the films was characterized by transmission electron microscopy (TEM, Tecnai G2 F20 S-Twin). The isothermal current-voltage ($I$-$V$) curves and the resistance versus temperature and magnetic field were measured using the standard four-probe method in a physical property measurement system (PPMS-6000, Quantum Design) equipped with $^3$He refrigerator. A Keithley 2400 Source Measure Unit was used to measure the $I$-$V$ curves. Hall-bar-shaped films (1.0-mm wide and 10.0-mm long, and the distance between the two voltage electrodes is 3\,mm), defined by using mechanical masks, were fabricated for the resistance and $I$-$V$ curve measurements. To obtain good contact, four Ti/Au electrodes were deposited on the films.

\begin{figure}[htp]
\includegraphics [scale=1.05]{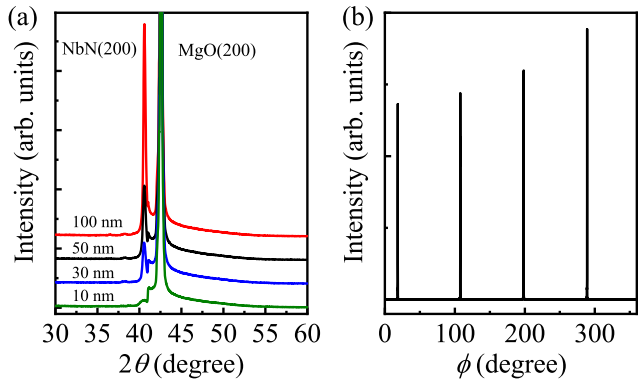}
\caption{(a) XRD $\theta$-2$\theta$ scan patterns of NbN films with different thickness. (b) $\phi$-scan spectrum of (220) plane for the 50-nm-thick NbN film.}\label{figXRD}
\end{figure}

\section{Results and Discussions}\label{SecIII}
\subsection{Crystal structure and morphologies}

Figure~\ref{figXRD}(a) shows the XRD $\theta$-$2\theta$ scan profiles for four NbN films with different thickness, as indicated. Besides the (200) diffraction peak of the MgO substrate, only the peak related to the (200) plane of $\delta$-NbN (Rocksalt structure) appears in each XRD pattern. Thus, $\delta$-NbN films without any impurity phase are successfully fabricated under the conditions mentioned above. The full width at half maximum of the (200) peak increases with decreasing film thickness, while the intensity of the peak decreases with decreasing thickness. When the thickness of the films is less than $\sim$10\,nm, the diffraction intensity of the film is too weak to be observed (the structure of these $t\lesssim 10$\,nm films will be further detected via HRTEM measurements). Figure~\ref{figXRD}(b) shows the $\phi$-scan profile of (220) plane of the 50-nm film (the $\phi$-scan profiles of those $t\gtrsim 30$\,nm films are similar).  Clearly, four uniformly distributed diffraction peaks appear in the profile, indicating that the NbN film is epitaxially grown on the MgO substrate.

\begin{figure}[htbp]
\includegraphics[scale=0.9]{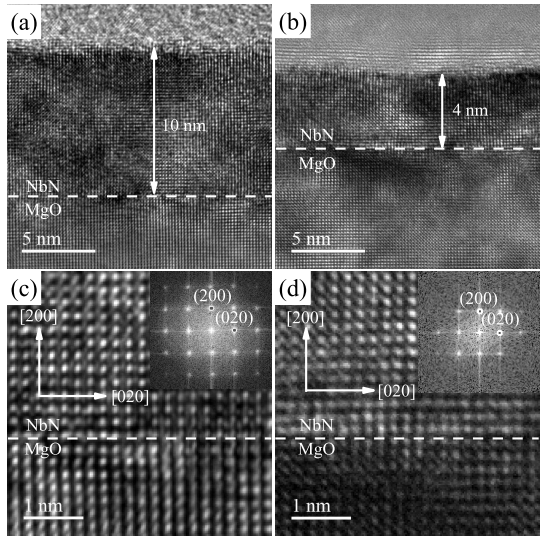}
\caption{Cross-sectional HRTEM micrographs of the (a) $t=10$\,nm and (b) $t=4.0$\,nm films. (c) and (d) are the enlarged images   near NbN/MgO interface for the two films, respectively. The insets in (c) and (d) are the fourier transform  of the HRTEM images of NbN film region for the corresponding films.}\label{figTEM}
\end{figure}

For the $t\lesssim 20$\,nm films, we use the TEM to detect their structures and morphologies. Figure~\ref{figTEM} shows the cross-sectional HRTEM micrographs of the 10.0- and 4.0-nm-thick films along the [001] axis of NbN, as indicated.  The NbN/MgO interface and the surface of NbN film can be clearly identified and the average thickness of each film can be obtained [see Fig.~\ref{figTEM}(a) and \ref{figTEM}(b)]. It is found that the thickness of each film obtained from TEM measurement is almost identical to that evaluated via growth rate and deposition time. Figure~\ref{figTEM}(c) and \ref{figTEM}(d) present the enlarged HRTEM images near the interface of NbN film and MgO substrate for the $t\simeq 10.0$ and 4.0\,nm films, respectively.  Combining the XRD result for the thick NbN films, one can readily obtain that the [200]-orientated $\delta$-NbN film is epitaxially grown on the (100) MgO substrate. The insets of Fig.~\ref{figTEM}(c) and \ref{figTEM}(d) are the fast Fourier transform images of the NbN films, which further confirm the epitaxy characteristics of the $\delta$-NbN films. We note in passing that the NbN film is still uniform and completely covers the MgO substrate even if the film is only as thin as $\sim$2.0\,nm.

\subsection{Berezinskii-Kosterlitz-Thouless transition}
\begin{figure}[htbp]
\includegraphics{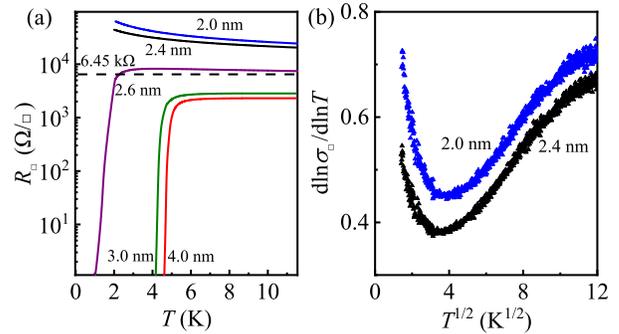}
\caption{(a) Sheet resistance $R_\square$ as a function of temperature $T$ for the NbN films with different thickness. (b)$\mathrm{d}\ln\sigma_\square/\mathrm{d}\ln T$ versus $T^{1/2}$ for the 2.4 and 2.0\,nm NbN films.}\label{figRT}
\end{figure}

Figure~\ref{figRT} (a) shows the sheet resistance $R_\square$ as a function of temperature for the NbN films with different thickness, as indicated. Clearly, those $t\gtrsim 2.6$\,nm NbN films reveal superconductor characteristics at low temperature regime, while those $t\lesssim 2.4$\,nm films exhibit insulator behaviors, i.e., $\mathrm{d} R_\square/\mathrm{d}T <0$ over the  whole measuring temperature range and $\mathrm{d}\ln\sigma_\square/\mathrm{d}\ln T\rightarrow\infty$ as $T\rightarrow 0$ [see Fig.~\ref{figRT}(b), here the sheet conductance $\sigma_\square$ is the reciprocal of $R_\square$]. It is worth noting that the behavior of the logarithmic derivative of the sheet conductance  $w=\mathrm{d}\ln\sigma_\square/\mathrm{d}\ln T$ near 0\,K is more reliable than the behavior of the derivative of the sheet resistance $\mathrm{d} R_\square/\mathrm{d}T$ in determining whether a certain sample is metallic or insulating~\cite{PrbMobius}. For the metallic sample, the value of $w$ tends to be zero as $T \rightarrow 0$, while $w$ tends to be a constant or divergent value as $T \rightarrow 0$ for the insulating sample. From Fig.~\ref{figRT} (a), one can also see that the NbN films directly change from superconductors to insulators  at low temperature with decreasing thickness and the transition (intermediate) states that often appear in granular films, such as ``quasireentrant" behavior or ``flat tail" of the low-temperature resistance ~\cite{PrbJaeger}, are not present.  In addition, the critical sheet resistance $R_\square^{\rm c}$ [the sheet resistance at normal state (10\,K)] for the SIT is very close to the quantum resistance of Cooper pairs $h/4e^2$ or 6.45\,k$\Omega$. According to Fisher~\cite{PrlFisher}, the critical sheet resistance for SIT in disordered 2D superconducting systems is universal, being close to the quantum resistance of Cooper pairs $h/4e^2$.

\begin{table}[htbp]
\caption{
Relevant parameters for the NbN films with different thickness $t$. Here $R_{\square}^{\rm N}$ is normal state sheet resistance at 10\,K, $T_{\rm BKT}$ is BKT transition temperature, $R_0$ and $b$ are the parameters in Halperin-Nelson formula, $C$ is the parameter in the activated scaling law (See the text), ${B_{\rm c}^\ast}$ is the characteristic critical field, and $T_{\rm M}$ is vortex melting temperature.}\label{tab:table2}
\begin{ruledtabular}
\begin{tabular}{cccccccc}
$t$   & $R_{\square}^{\rm N}$   & $T_{\rm BKT}$ & $R_0$              & $b$ & $C$ & $ B_{\rm c}^\ast$ & $T_{\rm M}$\\
 (nm) &(k$\Omega/\square$)  & (K)           & ($\Omega/\square$) &     &     &  (T)        & (K)\\
\hline
4.0 & 2.31 & 4.59 & 13479.50 & 0.71 & 0.40 & 8.58 & 5.78\\
3.0 & 2.84 & 4.21 & 15027.23 & 0.75 & 0.34 & 5.83 & 3.72\\
2.6 & 7.59 & 1.43 & 27115.80 & 1.09 & 0.29 & 3.87 & 0.83\\
\end{tabular}
\end{ruledtabular}
\end{table}

According to Koushik\emph{ et al.}~\cite{PrlKoushik}, the coherence length $\xi(0)$ of the NbN films is $\sim$6\,nm. Thus the NbN films with $t<6$\,nm are 2D with respect to superconductivity. A 2D superconductor with high normal state sheet resistance and lateral size of order of the transverse penetration depth $\lambda_\perp$ [$\lambda_\perp=\lambda^2(T)/d $ with $\lambda(T)$ being the usual bulk magnetic penetration depth] is expected to exhibit the Berezinskii-Kosterlitz-Thouless (BKT) transition~\cite{JetpBerezinskii, SspKosterlitz1972, SspKosterlitz1973, PrlVfolf}. Below the BKT transition temperature $T_{\textrm{BKT}}$ (which is somewhat less than the usual superconducting transition temperature $T_{\rm c}$), the vortices are bound in vortex-antivortex pairs. In each vortices pair, the helicities of the two vortices are opposite, and the separation (distance) between the two vortices is less than $\lambda_\perp$. Above $T_{\textrm{BKT}}$, the pairs are thermally dissociated and the motion of the dissociated vortex pairs should lead to a broadened superconducting transition.

\begin{figure}[htbp]
\includegraphics{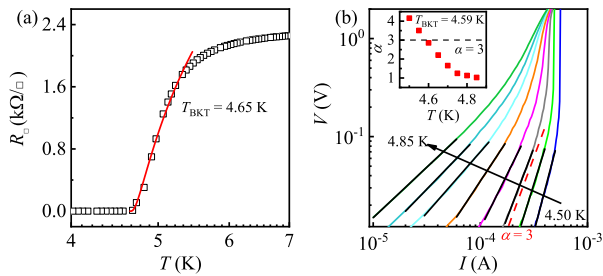}
\caption{(a) The sheet resistance $R_\square$ vs temperature $T$ for the 4.0-nm-thick film. The solid curve is the least-squares fits to Eq.~(\ref{Eq-Halperin-Nelson}). (b) Voltage vs current (double logarithmic scales) measured at fixed temperatures ranging from 4.50 to 4.85\,K at zero magnetic field. The temperature difference between two adjacent curves is 0.05 K. The solid black lines are least-squares fits to $V \sim I^{\alpha(T)}$, and the dashed line represents the line with $\alpha = 3$.  Inset: the exponent $\alpha$ vs temperature for the film.}\label{figBKT}
\end{figure}

Experimentally, the occurrence of BKT transition can be identified by analyzing the variations of the characteristic physical quantities near $T_{\rm BKT}$, such as, the temperature dependence of sheet resistance\cite{PrlDynes1978, PrlOvadyahu1980}, the current-voltage curves~\cite{PrlEpstein}, the \emph{rf} surface impedance~\cite{PrlHebard1990}, and the voltage noise spectra~\cite{PrlVoss}. For a 2D superconductor with BKT transition, the temperature dependent behavior of the sheet resistance can be described by the Halperin-Nelson formula~\cite{JltpHalperin}
\begin{equation}\label{Eq-Halperin-Nelson}
R_\square = R_0\textrm{exp}[-b(T/T_{\textrm{BKT}} - 1)^{-1/2}]
\end{equation}
as the transition temperature $T_{\rm BKT}$ is approached from above. Here both $R_0$ and $b$ are constants. Figure~\ref{figBKT}(a) shows the temperature dependence of the sheet resistance from 4 to 7\,K for the 4.0-nm-thick film. The solid curve is the least-squares fit to Eq.~(\ref{Eq-Halperin-Nelson}). In the fitting process, $R_0$ and $b$ are the adjustable parameters, and $T_{\rm BKT}$ is obtained by extrapolating the linear part of $[\mathrm{d}\ln R_\square /\mathrm{d}T]^{-2/3}$ vs $T$ curve to $[\mathrm{d}\ln R_\square /\mathrm{d}T]^{-2/3} = 0$.  Clearly, the $R_\square(T)$ data of the 4.0-nm-thick film near the transition region (above $\sim$4.65\,K) can be well described by Eq.~(\ref{Eq-Halperin-Nelson}). The $R_\square(T)$ data near the transition regions for the $t=3.0$ and 2.6\,nm films also obey the prediction of Eq.~(\ref{Eq-Halperin-Nelson}). For each film, the adjusting parameters $R_0$ and $b$, as well as the value of $T_{\rm BKT}$, are listed in Table~\ref{tab:table2}. Inspection of the table indicates that the values of $b$ for the three samples are all around $1$, which is consistent with the theoretical predication~\cite{JltpHalperin}.

As mentioned above, the occurrence of BKT transition can also be evaluated via the measurements of $I$-$V$ curves around the transition region. For the 2D superconductor with BKT transition, Kadin \emph{et al}.~\cite{PrbKadin}, have demonstrated that in the transition regime and in the limit of arbitrarily small current, the current dependence of voltage obeys $V \sim I^{\alpha(T)}$, where $\alpha(T)$ is a measure of the areal superelectron density and $\alpha(T) = 3$ at $T_{\textrm{BKT}}$. In Fig.~\ref{figBKT}(b), we present a set of $I$-$V$ curves for the $t \simeq 4.0$\,nm film. Clearly, $\log_{10} V$ varies linearly with $\log_{10}I$ in small current limit at each selected temperature, which means that the voltage variation with the current obeys the power law [$V \sim I^\alpha$]. The temperature dependence of $\alpha$ is given in the inset of Fig.~\ref{figBKT}(b). As the temperature is increased from 4.50 to 4.85\,K, the value of $\alpha$ decreases from $\sim$4.2 to $\sim$1.0, and the temperature for $\alpha=3$ is $\sim$4.59\,K, i.e., the value of $T_{\textrm{BKT}}$ is $\sim$4.59\,K, which is almost identical to that determined by the Halperin-Nelson formula. Thus the occurrence of BKT transition in the film  is confirmed. Similar characteristics are also observed in the $I$-$V$ curves of the $t=3.0$ and 2.6\,nm films, and the transition temperatures are also close to those determined by the Halperin-Nelson formula. Thus, these $t\lesssim 4.0$\,nm NbN films are 2D with respect to the superconductivity. We will focus on these three films in the following discussions.

\begin{figure}[htbp]
\includegraphics{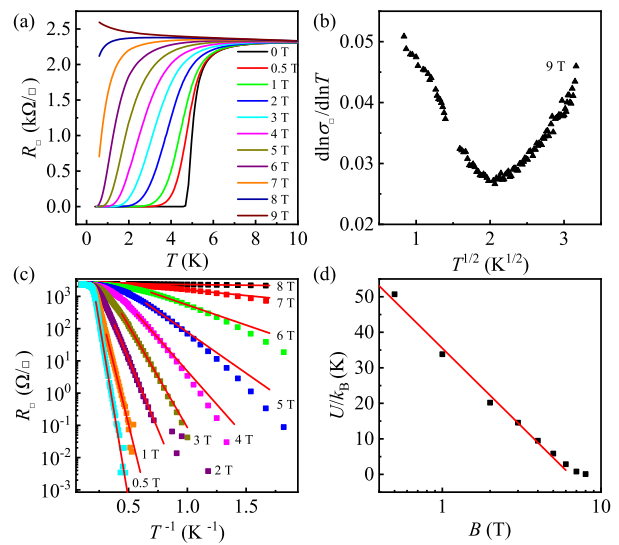}
\caption{(a) $R_\square (T)$ as a function of temperature for the 4.0-nm-thick NbN film measured at different fields perpendicular to the film plane. (b) $\mathrm{d}\ln\sigma_\square/\mathrm{d}\ln T$  vs $T^{1/2}$ for the 4.0-nm-thick NbN film under 9\,T. (c) Logarithm of the sheet resistance as a function of $T^{-1}$ at different magnetic fields for the 4.0-nm-thick film. The symbols are the experimental data and the straight solid lines are least-squares fits to Eq.~(\ref{Eq-ThermalA}). (d) Activation energy $U(B)/k_{\rm B}$, obtained from the slopes of the solid lines in (c), as a function of magnetic field. The solid line is the least-squares fit to  $U(B) = U_0 \ln (B_0/B)$.}\label{figQGS}
\end{figure}

\subsection{Quantum phase transition}
Considering the quantum phase transition can be induced by an external magnetic field, we measure the temperature dependence of $R_\square$ under different fields for the $t\simeq 4.0$, 3.0 and 2.6\,nm films. Since the results of the three films are similar, we only give and discuss those obtained from the $t\simeq4.0$ film. Figure~\ref{figQGS}(a) shows $R_\square$ as a function of $T$ from 10 to 0.5\,K under different fields. For $B\lesssim 5.0$\,T, the film is in superconducting state at low temperature. As the field is increased to 9.0\,T, the film converts into an insulator. Here the insulator still refers to the sample with $\mathrm{d} R_\square/\mathrm{d} T <0$ and $\mathrm{d}\ln\sigma_\square/\mathrm{d}\ln T|_{T\rightarrow 0}\rightarrow \infty$ or constant [see Fig.~\ref{figQGS}(b)]. The critical field of SIT lies between 8 and 9\,T. To check whether there is intermediate metal state in the film, the $R_{\square}(T)$ data in Fig.~\ref{figQGS}(a) are redrawn as $R_\square$ (in logarithmic scale) versus $T^{-1}$, shown in Fig.~\ref{figQGS}(c). Upon cooling, the film directly changes from normal to superconducting states as the external field is less than $\sim$5\,T. When the applied field is between $\sim$5 and $\sim$7\,T, the sheet resistance drops sharply below $T_{\rm c}(B)$, where $T_{\rm c}(B)$ is designated as the temperature at which the resistance drops to 90\% of the normal state resistance $R_{\square}^{\rm N}$ under $B$. The saturation trend appearing in the 2D superconductors with field-driven anomalous metal state was not observed. Inspection of Fig.~\ref{figQGS}(c) also indicates that the $\log_{10} R_\square$ (or $\ln R_\square$) varies linearly with $T^{-1}$ below $T_{\rm c}(B)$. Thus, the sheet resistance can be described by
\begin{equation}\label{Eq-ThermalA}
R_\square=R_0(B)\exp[-U(B)/k_{\rm B} T]
\end{equation}
around the transition region, where $R_0(B)$ is a prefactor, $k_{\rm B}$ is the Boltzman constant, and $U(B)$ is the activation energy under $B$. This means that the vortex-antivortex pairs are thermally dissociated into vortices and the motion of thermally activated individual vortices dominate the charge transport process in the transition region. The experimental $R_\square(T)$ data at a certain $B$ are least-squares fitted to Eq.~(\ref{Eq-ThermalA}) and the results are shown as the solid lines in Fig.~\ref{figQGS}(c). Thus, the activation energy $U(B)$ as a function of $B$ can be obtained and is shown in Fig.~\ref{figQGS}(d). Clearly, $U(B)$ varies linearly with $\log_{10}(1/B)$ ($\ln (1/B)$), i.e., $U(B)$ vs $B$ satisfies $U(B)=U_0\ln (B_0/B)$, which is in accordance with the theoretical predication of thermally assisted collective vortex-creep model for 2D superconductor~\cite{PcFeigel'Man}.

\begin{figure}[htbp]
\includegraphics{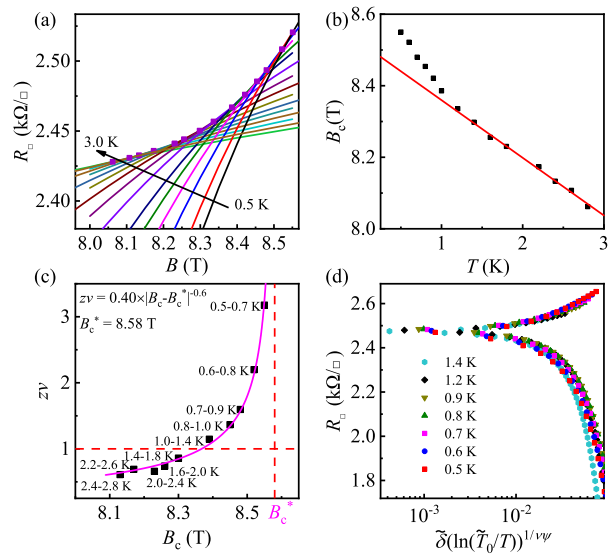}
\caption{(a) The sheet resistance vs magnetic field at different temperature for the 4.0-nm-thick NbN film. The temperature interval between two adjacent curves is 0.05\,K.  (b) Temperature dependence of crossing points (denoted as $B_{\rm c}$) of $R_\square(B)$ at every two adjacent temperature. The solid line is only the guide for eyes. (c) The critical field $B_{\rm c}$ dependence of critical exponent $z\nu$. (d) Sheet resistance vs the scaling parameter described in Eq.~(\ref{Eq-Activated2}) for activated scaling.}\label{figActivation}
\end{figure}

As mentioned in Sec.\ref{secI},  the ``dirty-boson model" predicts that the $R_\square$ vs $B$ curves at different temperatures all cross at the critical field $B_{\rm c}$. To check whether this model is suitable for the 2D NbN films, the magnetoresistance isotherms at temperatures from 0.5 to 3\,K have been measured and are shown in Fig.~\ref{figActivation}(a). Clearly, these $R_\square$-$B$ curves do not cross at one point, but the crossing points of two adjacent isotherms form a continuous curve over a range of temperatures and magnetic fields. This unusual phenomenon is very similar to that in three-monolayer Ga films~\cite{SciXing} and is the signature of the QGS. Designating the magnetic field at each crossing point as the critical field $B_{\rm c}$, one can readily obtain the critical field variation with temperature, which is shown in Fig.~\ref{figActivation}(b). Upon cooling, the critical field almost increases linearly with decreasing temperature from 3 to $\sim$1.2\,K, and then increases more rapidly below $\sim$1.2\,K. Assuming the $R_\square$-$B$ curves measured at three adjacent temperatures cross at one point, we rewrite Eq.~(\ref{Eq-scaling}) as $R_\square(B, t)=R_\square^{\rm c} f[(B-B_{\rm c})t]$ [with $t=(T/T_0)^{-1/z\nu}$ and $T_0$ being the lowest temperature] and use the equation to analyze the magnetoresistance isotherms: the $R_\square (B,T)$ data for the three adjacent temperatures would collapse onto two branches in the $R_\square (B,T)$ vs $\delta t$ plot as a suitable $t$ is selected. Thus the effective critical exponent $z\nu$ at a certain temperature range can be obtained, and $z\nu$ vs $B_{\rm c}$ for the 4-nm-thick film is summarized in Fig.~\ref{figQGS}(c). In the high temperature regime ($T\gtrsim 1.2$\,K), the effective critical exponent $z\nu$ increases slowly with increasing $B_{\rm c}$ , then increases dramatically in the low temperature regime, and finally tends to diverge as $B_{\rm c} \rightarrow B_{\rm c}^\ast$, where $B_{\rm c}^\ast$ is a characteristic field. The behavior of $z\nu$ is quite different to that of the conventional SIT in highly and homogeneously disordered 2D superconductors, in which $z\nu$ keeps as a constant in the vicinity of the transition. The $z\nu$ vs $B_{\rm c}$ data are least-squares fitted to the activated scaling law~\cite{PreVojta} $z\nu=C|B_{\rm c}-B_{\rm c}^\ast|^{-\nu\psi}$ with $C$ being a constant, the correlation length exponent $\nu=1.2$, and the tunneling exponent $\psi=0.5$, and the fitting result is shown by the solid curve in Fig.~\ref{figActivation}(c). Clearly, the experimental $z\nu$ vs $B_{\rm c}$ data can be well described by the activated scaling law with $B_{\rm c}^\ast\approx 8.58$\,T, which is the evidence of QGS associates with an infinite-randomness critical point.

On the other hand, the theory~\cite{PrlDelMaestro} also predicts that the sheet resistance vs magnetic field at different temperatures can be described by an activated scaling form
\begin{equation}\label{Eq-Activated2}
  R_\square\left( \tilde{\delta}, \ln\frac{\tilde{T}_0}{T} \right) = \Phi \left[ \tilde{\delta}\left(\ln\frac{\tilde{T}_0}{T}\right)^{1/\nu\psi}\right],
\end{equation}
where $\tilde{\delta}=|B-B_{\rm c0}|/B_{\rm c0}$ with $B_{\rm c0}$ being the critical field, and $\tilde{T}_0$ is a characteristic temperature. Taking $\nu\psi = 0.60$, we check whether the low-temperature $R_\square (T, B)$ data can be described by Eq.~(\ref{Eq-Activated2}). Fixing $B$ at a certain value of $B_{\rm c}$ shown in Fig.~\ref{figActivation}(b), we first get the critical field $B_{\rm c0}$ via minimizing the variance $\sum_{i=1}^{n-1} |R_\square(T_i, B)-R_\square(T_{i+1}, B)|^2$, where $T_i$ is a testing temperature. The obtained value of $B_{\rm c0}$ is 8.44\,T, which is 1.6\% less than the value of $B_{\rm c}^*$. Then $\tilde{T}_0$ is set as an adjustable parameter, and it is found that the $R_\square(T,B)$ vs $\tilde{\delta} [\ln(\tilde{T}_0/T)]^{1/\nu\psi}$ data almost collapse onto two branches as $\tilde{T}_0\approx 1.67$\,K. Thus the $R_\square(T,B)$ data obey the activated dynamical scaling law [Eq.~(\ref{Eq-Activated2})], which strongly suggests that the quantum superconductor-insulator transition in 2D NbN film is governed by an infinite-randomness fixed point with activated dynamical scaling from the other side. Inspection of Fig.~\ref{Eq-Activated2}(d) indicates that the $R_\square(T,B)$ data deviate from the scaling law for both the upper and low branches when the values of the scaling parameter (the value of the abscissa) is larger than $\sim$0.013. In this region, the upper branch corresponds to the field being higher than $\sim$8.6\,T, while the low branch corresponds to the field being less than $\sim$8.3\,T. In the high field (low field) case, the film has left the quantum critical regime and the influence of the quantum fluctuation on the properties of the film  is negligibly weak, which causes the $R_\square(T,B)$ data to deviate from the scaling rule of Eq.~(\ref{Eq-Activated2}) in the high scaling parameter regime.

Theoretically, the origin of the QGS is the quenched disorder~\cite{PreVojta, PrlDelMaestro}. In an epitaxial NbN film, the quenched disorder mainly caused by the intrinsic defects, such as nitrogen vacancies, Nb interstitial atoms, and dislocations. At low temperatures, the phase diagram of the  epitaxial NbN film is determined by the interplay of the quantum fluctuation, quenched disorder, and thermal fluctuation. As mentioned above, the critical field $B_{\rm c}$ increases abruptly below $\sim$\,1.2\,K. Inspection of Fig.~\ref{Eq-Activated2}(c) indicates that the $z\nu$ value also tends to be greater than 1 below $\sim$1.2\,K. It is generally considered that $z\nu >1$ is a hallmark of the occurrence of QGS, thus the abrupt increase of $B_{\rm c}$ below $\sim$1.2\,K is caused by the quenched disorder. The temperature for $z\nu =1$ is defined as the vortex melting temperature $T_{\rm M}$\cite{SbHuang}. Below $T_{\rm M}$, the quantum fluctuation dominates over the thermal fluctuation, on the other side the influence of quenched disorder on the vortices also becomes powerful. Finally, at high field [but $B\lesssim B_{\rm c}(T)$] and $T< T_{\rm M}$ regime, the system is transformed into a vortex glass-like phase composed of spatially separated superconducting rare regions (puddles or islands). These sizes of these rare regions increase with decreasing temperature, and the slow dynamics results in the divergency of the critical exponent $z\nu$. Recently, the high field vortex glass-like state~\cite{EplKundu} and the spatial inhomogeneity of superconducting regions~\cite{PrbNoat, PrbCarbillet2013, PrbCarbillet2016} in quasi-2D NbN superconducting films had been observed experimentally, which strongly supports the scenario of the origin of QGS mentioned above.

The divergence of $z\nu$ in SIT is also observed in the $\sim$3.0 and $\sim$2.6\,nm NbN films. For these two films, the critical exponent $z\nu$ vs the critical field $B_c$ also obeys the activated scaling law $z\nu=C|B_{\rm c}-{B_{\rm c}^\ast}|^{-0.6}$. The values of $C$ and $B_{\rm c}^{\ast}$ for the two films are listed in Table~\ref{tab:table2}. Table~\ref{tab:table2} also gives the normal state sheet resistance $R_{\square}^{\rm N}$ (the sheet resistance at 10\,K) and vortex melting temperature $T_{\rm M}$ for the $\sim$4.0, $\sim$3.0, and $\sim$2.6\,nm films.  One can see that the vortex melting temperature $T_{\rm M}$ decreases with decreasing $R_{\square}^{\rm N}$. Considering that $R_{\square}^{\rm N}$ represents the disorder degree of a 2D superconducting system, one can readily obtain that the vortex melting temperature $T_{\rm M}$ decreases with increasing disorder degree. However, the value of $T_{\rm M}$ is still as high as $\sim$0.82\,K even for the most disordered film (the 2.6-nm-thick film).

\begin{figure}[htp]
\includegraphics{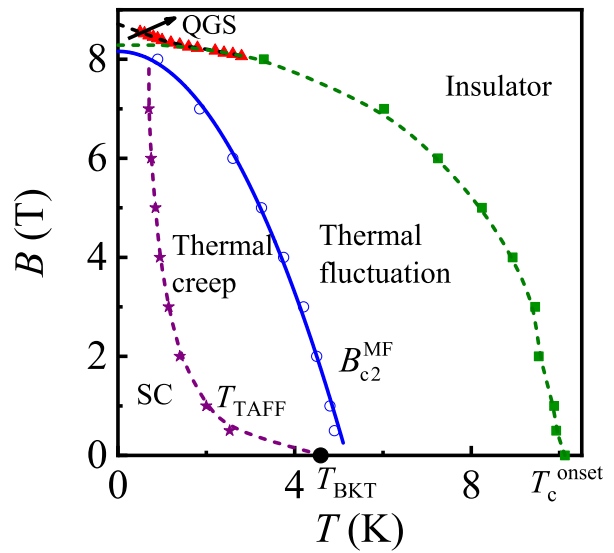}
\caption{$B$-$T$ phase diagram of the 4.0-nm-thick NbN film. The triangles, squares, hollow circles, stars represent the crossing field $B_\textrm{c}(T)$, $T_\textrm{c}^{\textrm{on}}(B)$, the mean-field critical field $B_\textrm{c2}^{\textrm{MF}}(T)$, and $T_{\textrm{TA}}$, respectively. The dashed curves are only the guides to eyes. The solid curve is least-squares fits to $B_{\rm c2}^{\rm MF}(T)/B_{\rm c2}^{\rm MF}(0)=1-({T}/{T_{\rm c0}})^2$.}\label{Fig-phaseDiagram}
\end{figure}

Finally, we give the low-temperature $B$-$T$ phase diagram of the 4.0-nm-thick NbN film as an example (Fig.~\ref{Fig-phaseDiagram}). In Fig.~\ref{Fig-phaseDiagram} the superconducting onset temperature $T_{\rm c}^{\rm on}(B)$ was obtained from the $R_\square$-$T$ curves at zero and finite magnetic fields [Fig.~\ref{figRT}(a)], and defined as the temperature for $\mathrm{d} R_\square/\mathrm{d}T = 0$ near the superconducting transition region. As the temperature is lower than $T_{\rm c}^{\rm on}(B)$ under $B$, the film falls into the thermal fluctuation region. In this region, $R_\square(B)$ starts to decrease with decreasing temperature due to the formation of Cooper pairs and the appearance of superconducting fluctuations. It should be noted that the $T_{\rm c}^{\rm on}(B)$ and the high field $B_{\rm c}(T)$ almost follow the same trajectory. Below $\sim$1.2\,K, the trajectory represents the boundary between insulator and quantum Griffiths state. The mean field upper critical field $B_{\rm c2}^{\rm MF}(T)$ is obtained according to the Ulah-Dorsey (UD) scaling theory~\cite{PrlUllah}. The calculation process is identical to that used in Ref.~[\onlinecite{NcSaito}]. The solid curve in Fig.~\ref{Fig-phaseDiagram} is the least-squares fits to the empirically formula~\cite{SSP-Ashcroft}, $B_{\rm c2}^{\rm MF}(T)=B_{\rm c2}^{\rm MF}(0)[1-({T}/{T_{\rm c0}})^2]$. Here the zero temperature critical field $B_{\rm c2}^{\rm MF}(0)$ and zero field critical temperature $T_{\rm c0}$ are taken as 8.16\,T and 5.18\,K, respectively. The temperature $T_{\rm TA}(B)$ is the characteristic temperature above which thermal assisted vortex creep governs the electrical transport of the film. The value of $T_{\rm TA}(B)$ is obtained from the $\log_{10}R_\square$ ($\ln R_\square$) vs $T^{-1}$ under different $B$ plot, which would depart from the linear part below $T_{\rm TA}(B)$ (the film gradually reaches superconducting state below $T_{\rm TA}(B)$). When a moderate magnetic field $B$ is applied, thermal assisted vortex creep is the dominant transport mechanism between $T_{\rm TA}(B)$ and $T_{\rm c}^{\rm MF}(B)$, where $T_{\rm c}^{\rm MF}(B)$ is the point in the $B_{\rm c2}^{\rm MF}$ vs $T$ curve. The QGS behavior would appear below $\sim$1.2\,K at relative high magnetic field due to the effect of the quenched disorder.

\section{Conclusion}
The NbN epitaxial films with different thickness  were grown on (100) MgO single crystal substrates by reactive magnetron sputtering method. The low-temperature electrical transport properties of these 2.0- to 4.0-nm-thick films have been systematically investigated. The normal-state sheet resistance (for example, the sheet resistance at 10\,K) increases with decreasing film thickness. When the normal-state sheet resistance excesses the quantum resistance of Cooper pairs $h/4e^2$, the ground state of the films changes from superconducting to insulating states. These superconducting films ($t\gtrsim 2.6$\,nm) are 2D with respect to superconductivity and undergo a BKT transition as transforming from normal to superconducing states upon cooling. For each 2D NbN superconducting film, a SIT can also be induced by a magnetic field perpendicular to the film plane.
Upon the field-driven transition process, the intermediate anomalous metal state does not present. However, the low-temperature magnetoresistance isotherms do not cross at a single fixed point but at a well-distinguished region. The critical dynamical critical exponent $z\nu$ tends to be divergent as $T\rightarrow 0$ and $B\rightarrow B_{\rm c}^{\ast}$, which suggests the occurrence of QGS behavior during the quantum phase transition. The QGS behavior arises from the low-temperature quenched disorder and is the consequence of the formation of the superconducting rare regions. Our results suggest that the QGS not only occurs in the highly crystalline 2D superconductors with superconductor-metal transition but also in those with SIT.

\begin{acknowledgments}
This work is supported by the National Natural Science Foundation of China through Grants No. 12174282 (Z.Q.L.) and No. 12074056 (P.L.).
\end{acknowledgments}

\end{document}